IAC-10.A3.6.5

# ODYSSEY 2: A MISSION TOWARD NEPTUNE AND TRITON TO TEST GENERAL RELATIVITY

**B. Lenoir**, B. Christophe, A. Lévy, B. Foulon
Office National d'Études et de Recherches Aérospatiales (ONERA), France, bruno.christophe@onera.fr

S. Reynaud, J.M. Courty, B. Lamine
Laboratoire Kastler Brossel (LKB), UPMC, CNRS, ENS, Paris, France

H. Dittus, T. van Zoest
DLR, Institute of Space System, Bremen, Germany

C. Lämmerzahl, H. Selig
ZARM, University of Bremen, Bremen, Germany

S. Léon-Hirtz, R. Biancale
Centre National d'Études Spatiales, Paris, France

G. Métris
Observatoire de la Côte d'Azur, Grasse, France

F. Sohl
DLR, Institute for Planetary Research, Berlin, Germany

P. Wolf
LNE-SYRTE, Observatoire de Paris, UPMC, CNRS, Paris, France

Odyssey 2 will be proposed in December 2010 for the next call of M3 missions for Cosmic Vision 2015-2025. This mission, under a Phase 0 study performed by CNES, will aim at Neptune and Triton. Two sets of objectives will be pursued.

The first one is to perform a set of gravitation experiments at the Solar System scale. Experimental tests of gravitation have always shown good agreement with General Relativity. There are however drivers to continue testing General Relativity, and to do so at the largest possible scales. From a theoretical point of view, Einstein's theory of gravitation shows inconsistencies with a quantum description of Nature and unified theories predict deviations from General Relativity. From an observational point of view, as long as dark matter and dark energy are not observed through other means than their gravitational effects, they can be considered as a manifestation of a modification of General Relativity at cosmic scales. The scientific objectives are to: (i) test the gravitation law at the Solar System scale; (ii) measure the Eddington parameter; and (iii) investigate the navigation anomalies during fly-bys.

To fulfil these objectives, the following components are to be on board the spacecraft: (i) the Gravity Advanced Package (GAP), which is an electrostatic accelerometer, based on ONERA's experience, to which a rotating stage is added; (ii) radio–science for precise range and Doppler measurement, and (iii) laser ranging, to improve significantly the measure of the Eddington parameter.

The second set of objectives is to enhance our knowledge of Neptune and Triton. Several instruments dedicated to planetology are foreseen: camera, spectrometer, dust and particle detectors, and magnetometer. Depending on the ones kept, the mission could provide information on the gravity field, the atmosphere and the magnetosphere of the two bodies as well as on the surface geology of Triton and on the nature of the planetary rings around Neptune.

## I. INTRODUCTION

Odyssey 2 is a space mission which combines fundamental physics experiments with planetary objectives, as recommended by the Fundamental Physics Roadmap Advisory Team appointed by ESA[1]. For this purpose, it will head toward the outer solar system: this will give the opportunity to test gravitation at the Solar System scale and visit Neptune and Triton.

Testing gravitation at the largest accessible scales is in line with the challenges currently facing fundamental physics. Indeed, even if General Relativity has met every experimental test up to now[2], there exists a set of yet unexplained possible anomalies[3]. A major concern is about the rotation curves of galaxies and the relation between red-shifts and luminosities of supernovae. These observations have been explained by introducing "dark components" in the Universe while using General Relativity. Dark matter is thought[4] to constitute approximately 25% of the energy content of the Universe, dark energy 70% and ordinary matter only 5%. However, despite their prevalence, they have never been observed by other means than gravitational ones. As a result, one option is to consider that General Relativity fails to describe physics at these scales[5]. It is therefore essential to test gravitation at the largest accessible scale. The Pioneer 10 and 11 missions performed such a test which resulted in what is now known as the Pioneer anomaly[6,7]. Because of the challenge raised by the Pioneer signal, the data have been reanalysed several times with the aim of understanding their origin: experimental artefact or hint of a new physics, currently investigated by theorists aiming at merging General Relativity with the description of the three other fundamental interactions. Some theories which are candidates for achieving such a unification – e.g. Scalar-tensor-vector gravity theory[8], generalized metric extensions of General Relativity[9] – may lead to violations of basics principles: (i) violation of the Lorentz invariance, (ii) violation of the equivalence principle and (iii) modification of the law of gravitation at microscopic or cosmological scales. Detecting this third violation will be one of the objectives of Odyssey 2 mission.

The cruise phase will also be used to test General Relativity through a high-accuracy measurement of the Eddington parameter $\gamma$ during solar conjunction. In the General Relativity framework, $\gamma$ is equals to one but some new theories predict a deviation from this value. The objective of Odyssey 2 mission is to gain a factor 200 with respect to the precision obtained by Cassini.

Before the cruise phase toward the outer Solar System, and depending on the strategy adopted, the spacecraft may experience one or more fly-bys of the Earth. These fly-bys, which aim at keeping the cost of the mission low by gaining energy from the planets, are an ideal opportunity to get more data concerning anomalies which have been observed on several recent Earth fly-bys[10].

These three fundamental physics objectives were the core of Odyssey mission[11] which was submitted by a large international team to ESA in response to the Cosmic Vision 2007 call. Through the proposal was not selected, its fundamental physics objectives are supported by the Fundamental Physics Roadmap Advisory Team which recommended to add planetary objectives.

Odyssey 2 will embark, in addition to the instruments dedicated to fundamental physics, instrumentation to study Neptune and Triton. Since this is an evolution from Odyssey mission, this instrumentation is still under discussion with the planetary community.

Odyssey 2, which is currently under a phase 0 study, will be proposed to ESA in response to the Cosmic Vision 2010 call in December 2010. The mission must fit in a cost frame of 470 M€, which includes the spacecraft, the launch and the operations: the instruments costs are supported by the national agencies.

## II. SCIENTIFIC OBJECTIVES

As mentioned in the introduction, Odyssey 2 will combine fundamental physics and planetary objectives. The measurement concerning fundamental physics are carried out during the interplanetary cruise whereas the one concerning planetology are carried out when the spacecraft fly-bys or orbits a body.

### II.I Fundamental physics scientific objectives

One of the key objectives of Odyssey 2 is to perform a comprehensive set of gravity tests in the Solar System. The mission has three objectives in the field of fundamental physics: (i) significantly improve the accuracy of deep space navigation; (ii) improve the accuracy of the measurement of the Eddington parameter; and (iii) investigate planetary fly-bys.

### Deep Space Gravitation

All orbit determination for interplanetary spacecraft are made using the Doppler acceleration along the line of sight, measured using the Doppler shift of the radio link[12]. This Doppler acceleration contains the total acceleration of the spacecraft and the effect of gravitation on light propagation. The total acceleration itself is the sum of the gravitational acceleration and the non-gravitational ones. As a result, in the orbit determination process, it is necessary to have information on the non-gravitational accelerations: currently models are used to correct for them. But these models induce uncertainty because of their inaccuracy or their inability to capture the real physical phenomenon. Figure 1, for example, shows the power spectrum density of this acceleration noise due to direct solar radiation pressure at 10 AU. This noise is impossible to predict and may only be captured by introducing additional parameters to be fitted during the orbit determination, which degrades the orbit determination process.

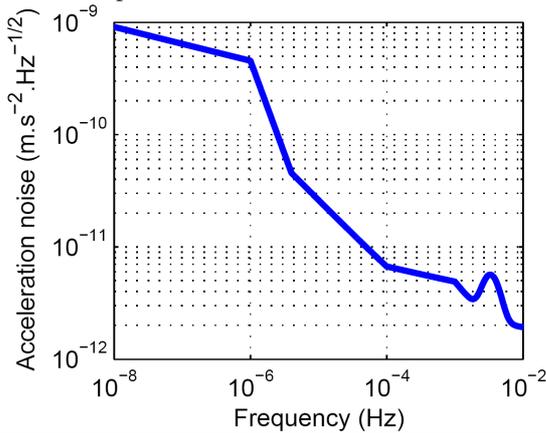

Fig. 1: Characterization (square root of the power spectrum density) of the solar radiation pressure fluctuations in term of acceleration of the spacecraft at 10 AU. The plot uses data from Fröhlich & Lean[13] and a ballistic coefficient for the spacecraft equal to $C_B = 0.1$ m$^2$.kg$^{-1}$. The bump at 3 mHz corresponds to the 5-minute oscillations of the Sun.

A key novelty of the Odyssey 2 mission is the presence of an accelerometer on-board, measuring directly the non-gravitational acceleration of the spacecraft. It provides an additional observable which therefore enhances the orbit reconstruction because it removes parameters to be fitted in the process, it therefore removes correlations and it also measures the fluctuations of the non-gravitational acceleration, which cannot be modelled.

This acceleration has different causes. First, the solar radiation pressure, already mentioned, corresponds to the kinetic energy transferred to the spacecraft by solar photons. The anisotropic thermal radiation of the spacecraft also induces a force, which is difficult to model because of the difficulty to know the evolution in time of the thermo-optics coefficients of the materials on the spacecraft[14,15,16]. During fly-bys or orbit of a body, the atmosphere induces a drag on the spacecraft and the spacecraft may also experience the radiation pressure due to the planet's albedo. Other forces, such as magnetic forces, may also play a role.

Given the previous considerations, the target precision in term of acceleration on the orbit determination is 10 pm.s$^{-2}$ along the line of sight. The observable quantities that will be used to achieve such a precision are:
- The range and the range rate between the spacecraft and the ground stations, removing if necessary the effects of the plasma by means of a multifrequency link. (cf. section IV.II)
- The non-gravitational accelerations acting on the spacecraft, measured by the accelerometer. (cf. section IV.I)
- The attitude of the spacecraft, measured by star trackers, which are commonly used on all spacecrafts for attitude determination.

Eddington parameter γ

The Eddington parameter γ, whose value in General Relativity is unity, is a key parameter in most test of gravitation. Indeed, the value (1-γ) is a mean to measure deviations from General Relativity of competing theories. The Cassini experiment[17,18] showed that this value is smaller than 2 x 10$^{-5}$. However, there exist theoretical models which suggest that (1-γ) has a value smaller than this experimental limit[19,20,21,22,23].

This is a motivation for repeating the Cassini experiment. This experiment is based on the measure, during a solar conjunction, of the Shapiro time delay, which is a purely relativistic effect and whose value is dependant on the Eddington parameter. The increase $\Delta t$ produced by the gravitational field of the Sun (with a mass $M$) in the time taken for light to travel the round trip between the ground station and the spacecraft is[24]

$$\Delta t = 2(1+\gamma)\frac{GM}{c^3}\ln\left(\frac{r_E + r_S + R}{r_E + r_S - R}\right), \qquad [1]$$

where $G$ is the gravitational constant, $c$ is the speed of light in vacuum, $r_E$ is the Earth–Sun distance, $r_S$ is the spacecraft–Sun distance and $R$ is the Earth–spacecraft distance.

Odyssey 2 will have in addition the advantage of the on-board accelerometer: since the non-gravitational acceleration was not measured by Cassini, it was fitted in the processing of the data, reducing the accuracy on γ.

The accuracy target on the Eddington parameter for the Odyssey 2 mission is 10$^{-7}$. It relies on the laser ranging equipment on-board Odyssey 2 (cf. section IV.III). This improvement in the precision comes from the fact that laser frequencies are much larger than radio frequencies. Indeed, the electron plasma around the Sun induces a time delay proportional to $f^{-2}$, where $f$ is the frequency of the link. Even if this delay can be corrected with dual-band radio link, the precision remains lower than with optical frequencies.

### Investigation of the fly-by anomaly

Several missions exhibit the following feature called generically fly-by anomaly[25]: the observed difference between the incoming speed and the outgoing speed, called $\Delta V$, is different from the one computed from precisely measured initial conditions and the well known property of the Earth gravity field. The discrepancies between the observed and computed $\Delta V$ are of the order of $10^{-2}$ m.s$^{-1}$, which is significantly larger than the measurement accuracies: the knowledge of Earth gravity field, atmospheric drag, charging and Earth tide has been shown to result in uncertainties well below the measured effect.

Odyssey 2 mission will use the opportunity of planetary fly-bys to investigate this anomaly with the advantage of the on-board accelerometer, as well as accurate tracking technique. The target precision on $\Delta V$ is 2 x $10^{-5}$ m.s$^{-1}$, with accelerometric measurement available during the whole duration of the fly-by.

## II.II Planetary scientific objectives

In addition to the fundamental physics objectives discussed above, Odyssey 2 will perform observations of Neptune and Triton. The planetary scientific objectives are still under discussion with the planetary community but the following topics may be addressed by the mission:
- Measure of the density profile and of the composition of the atmospheres;
- Retrieval of the gravitational fields of Neptune and Triton in order to better understand their interior structure;
- Imaging of the surfaces and measurement of their composition;
- Measure of the magnetospheres.

## III. DESCRIPTION OF THE ODYSSEY 2 MISSION

### III.I Preliminary mission analysis

The nominal launch date for an M3 mission is 2022. Considering this, two mission profiles have been studied.

The first one considers a launch with Soyouz. This would cost 75 M€, to be compared 470 M€ which is the overall cost of an M-mission (without the instruments). Such a launcher is not able to provide enough energy to the spacecraft so that it is able to reach the outer Solar System. As a result, trajectories with several fly-bys are envisioned: either a Venus-Earth-Earth trajectory or a Venus-Venus-Earth trajectory. In term of duration, the spacecraft would spend 3 years in the inner Solar System and it would need 14 to 15 years to go to Neptune after the last Earth fly-by. This would lead to an arrival at Neptune in 2039 or 2040.

The other mission profile is designed with a direct launch using Ariane, which costs 175 M€. Directly after launch, the spacecraft would head directly to Neptune, with no gravity assist manoeuvres. In this case, the spacecraft would reach Neptune in 14 or 15 years, leading to an arrival date in 2036 or 2037. This mission profile requires however to drop the planetary fly-by scientific objective.

The mission durations are relatively long because no gravity assist manoeuvres are possible with Jupiter or Saturn for a launch date between 2020 and 2029. There is, however, in the ESA schedule a fast-track option for the M3 mission in case of a delay for the L1 mission. This fast track option consists in a launch in 2020, which allows for Jupiter and/or Saturn gravity assist, reducing the duration of the mission to ten years leading to an arrival at Neptune in 2028 or 2029. Such an opportunity is again available in 2029.

### III.II Technical considerations

Because the spacecraft will head towards the outer Solar System, power generation is a concern. Indeed, solar panels cannot be considered: the size needed to power the spacecraft as well as all the instruments during the planetary phase would be too large.

As a result, nuclear generation is the only possible source of energy for such a mission. Two options can be considered. The first one is to build a US collaboration since Radioisotope Thermoelectric Generators (RTG) are readily available. More precisely, collaboration with ARGO mission[26,27] may be envisioned. The other option is for Europe to develop the use of nuclear energy in space.

The other major technical issue on the spacecraft design concerns self-gravity. The GAP instrument (discussed in the next section) is sensitive to the self-gravity of the spacecraft. As a result, it has to be monitored carefully.

The main source of uncertainty on self-gravity is the ergols: The uncertainties on the remaining quantity and on the position of the ergols in the tanks are directly impacting the knowledge of the self-gravity of the spacecraft. To control this source of uncertainty on the measurement, several tanks are used and the accelerometer is located between them.

## IV. INSTRUMENTS

IV.I An electrostatic accelerometer with bias rejection: GAP

The electrostatic accelerometer with bias rejection[28], is composed of an electrostatic accelerometer[29], called MicroSTAR, to which a rotating stage, called Bias Rejection System, has been added. Indeed MicroSTAR suffers a bias and is therefore able to make only differential measurement. Since one of the goals of the mission requires to measure absolute accelerations, the Bias Rejection System is added to be able to calibrate the bias and remove it. In addition to these two subsystems, an interface and Control Unit (ICU) is in charge of controlling the instrument. The target consumption, mass and volume of the whole instrument are 3 W, 3.1 kg and 3 L.

MicroSTAR is based on ONERA's expertise in the field of accelerometry and gravimetry: CHAMP, GRACE and GOCE missions[30,31] and the upcoming Microscope mission[32,33,34]. Therefore, MicroSTAR uses technology that already flew in space with improvement to meet the requirement in term of mass, consumption and volume. The principle of the accelerometer is to measure and control the motion of the proof mass by capacitive detection. Figure 2 shows the core of MicroSTAR with the proof mass and the electrodes.

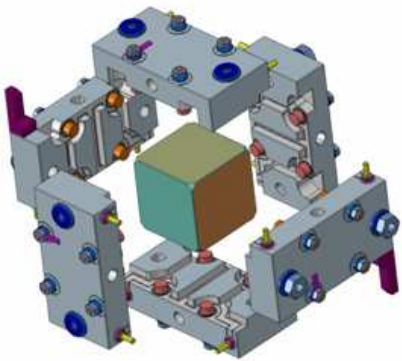

Fig. 2: Exploded view of the mechanical core of MicroSTAR. The position and attitude of the proof mass is controlled by six pairs of electrodes whose potentials are defined by a control loop. The potentials on the proof mass ±Vp are kept precisely constant by two gold wires. On each ULE plates, four stops prevent the proof mass to touch the electrodes.

Models, ground experiments and in-orbit measurements allow characterizing MicroSTAR's performances in term of noise. Figure 3 shows, for a measurement range of $1.8 \times 10^{-4}$ m.s$^{-2}$, the square-root of the power spectrum density of the noise.

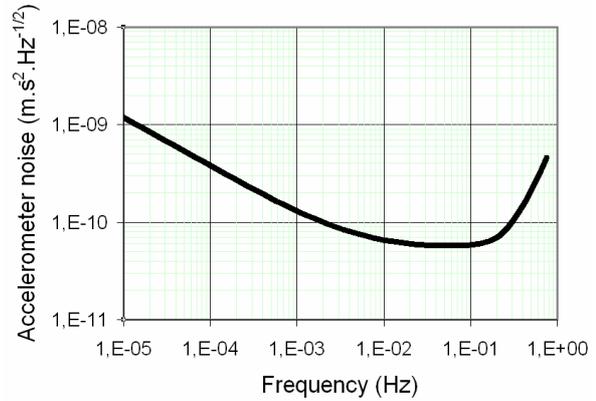

Fig. 3: Square-root of the Power Spectrum Density of MicroSTAR's noise.

Assuming that the Bias Rejection System rotates the accelerometer with a period of 10 minutes and that the integration time is four hours, the precision achieved on the measured acceleration is $10^{-12}$ m.s$^{-2}$. This value is below the requirement on deep space navigation which has a target precision of $10^{-11}$ m.s$^{-2}$.

IV.II Multi-frequency radio links

Radio link is used for data transmission as well as orbit determination. When using a one-frequency radio link, due to the dispersion of electromagnetic waves in plasmas, the Doppler observable incorporates the plasma effect which cannot be removed. It induces a time delay $T$ given by

$$T = \frac{\mu_0 c^2 e^2}{8 \, m_e \pi^2} \frac{\int N_e dl}{f^2 c} \approx \frac{40.3 \ m^3.s^{-2}}{f^2 c} \int N_e dl, \text{[2]}$$

where $\mu_0$ is the vacuum permeability, $e$ and $m_e$ are respectively the charge and mass of the electron, $f$ is the frequency of the electromagnetic wave, $N_e$ is the electron density in the plasma and $\int N_e dl$ is the integral over the optical path.

Multi-frequency radio links takes advantage of the dispersion in plasma to calibrate the integral over the electron density and remove its contribution. This technique has been used on Cassini for the measurement of the Eddington parameter and it will be used on Bepi-Colombo mission[35].

This instrument will be necessary to achieve the aimed precision for deep space navigation. For the measure of the Eddington parameter, laser ranging, presented below, will be used.

IV.III Laser ranging

Considering the time delay discussed in equation [2], one sees that it is negligible for optical frequencies. Therefore, for the measure of the Eddington parameter during solar conjunction, laser ranging will be used in order to have a measure not perturbed by the solar plasma. The use of laser ranging requires the satellite and the ground stations to be equipped with optical telescopes.

IV.IV Planetary instrumentation

In addition to the instrumentation discussed above, Odyssey 2 will embark several instruments for the study of Neptune and Triton. These instruments are currently under discussion with the planetary community. The mission should incorporate a camera to image the surface. The determination of the composition of the environment will require having a spectrometer and a particle detector on board. Finally, a magnetometer will be needed to map the magnetosphere around Neptune and Triton.

In addition to these dedicated instruments, the accelerometer described previously will be used during the planetary phase of the mission. Indeed, during orbits or fly-bys, the spacecraft experiences non-gravitational forces due for example to the atmosphere. The orbit reconstruction around a planet allows deriving the gravity field of this planet. The precision on the orbit reconstruction therefore directly impact the quality of the computed gravity field. By producing an additional observable, the accelerometer will, as for deep space navigation, allow for a higher quality of the scientific return for the planetary phase of the mission.

V. CONCLUSION

The Odyssey 2 mission relies on several European institutes specialized in fundamental physics, precise navigation instrumentation, accurate orbit determination or planetology. It will be proposed to the Cosmic Vision 2010 call with the advantage of combining fundamental physics experiments with planetary objectives on Neptune and Triton.